\documentclass[conference]{IEEEtran}

\usepackage{setspace}
\onecolumn

\newcommand\blfootnote[1]{%
  \begingroup
  \renewcommand\thefootnote{}\footnote{#1}%
  \addtocounter{footnote}{-1}%
  \endgroup
}

\usepackage{graphicx}
\makeatletter
\def\Ginclude@eps#1{%
 \message{<#1>}%
  \bgroup
  \def\@tempa{!}%
  \dimen@\Gin@req@width
  \dimen@ii.1bp%
  \divide\dimen@\dimen@ii
  \@tempdima\Gin@req@height
  \divide\@tempdima\dimen@ii
    \includegraphics{#1}%
  \egroup}
\makeatother

\usepackage[square, comma, sort&compress, numbers]{natbib}
\usepackage{flushend}

\usepackage[table]{xcolor}  
\usepackage[bookmarks=true,pdfborder={0 0 0}]{hyperref}

\usepackage{amsmath}
\usepackage{algorithmic}
\usepackage{algorithm}
\begin{document}
\onehalfspacing
\bstctlcite{IEEEexample:BSTcontrol}
\title{A Technique for Efficiently Managing SRAM-NVM Hybrid Cache}

 \author{\IEEEauthorblockN{Sparsh Mittal }
 \IEEEauthorblockA{Department of Electrical and Computer Engineering \\
 Iowa State University, Ames, Iowa 50011, USA\\
  Email: sparsh0mittal@gmail.com}
 }

\maketitle

\begin{abstract}
In this paper, we present a SRAM-PCM hybrid cache design, along with a cache replacement policy, named dead fast block (DFB) to manage the hybrid cache. This design aims to leverage the best features of both SRAM and PCM devices. Compared to a PCM-only cache, the hybrid cache with DFB policy provides superior results on all relevant evaluation metrics, viz. cache lifetime, performance and energy efficiency. Also, use of DFB policy for managing the hybrid cache provides better results compared to LRU replacement policy on all the evaluation metrics. 
\end{abstract}

\begin{IEEEkeywords}
Non-volatile memory (NVM), SRAM-NVM Hybrid Cache, Phase Change Memory, Write Endurance, Cache Replacement Policy, Energy Efficiency
\end{IEEEkeywords}

\IEEEpeerreviewmaketitle

\section{Introduction}

With increasing performance-demand placed on state-of-the-art applications \cite{raju2012high,khaitan2013highchapter} and increasing system core-count, the pressure on memory system has increased. Since memory bandwidth increases at much slower rate than the core-count, processor designers have recently turned towards using very large last level caches (LLCs). Conventionally, SRAM has been used to design on-chip caches. However, SRAM has high leakage power consumption and hence, large LLCs designed with SRAM consume a significant fraction of processor power. 

Recently, researchers have explored used of NVM (non-volatile memory) devices.  NVMs offer high density, low leakage power and better scalability compared to the SRAM. At the same time, NVMs are not strictly superior to SRAM on all design parameters. Specifically, PCM has very high write latency/energy and very low write endurance \cite{mittalPCMsurey2013}. 

To address these issues, we propose a way-based SRAM-PCM hybrid cache design where 2 out of 8 ways are designed using SRAM and the remaining ways are designed using PCM. Our hybrid cache design aims to leverage the fast access speed and high write endurance of SRAM and the low-leakage power and high density of PCM. To increase the number of writes to the SRAM ways, we also propose DFB replacement policy, which aims to evict dead SRAM (fast) ways much before they reach the bottom of the LRU stack. \blfootnote{Sparsh Mittal is currently working as a postdoctoral research associate at Oak Ridge National Laboratory, USA.}

We perform out-of-order simulations using Sniper x86-64 simulator \cite{CarHei2011_Sniper} and benchmarks from SPEC2006 suite. The results show that, compared to a PCM-only cache, our hybrid cache design with DFB replacement policy provides 6.9$\times$ improvement in cache lifetime (with as high as 13736$\times$ for povray). Also, it provides 4.6\% saving in memory subsystem energy and improves the performance by 1.36$\times$. In contrast, the hybrid cache with LRU replacement policy provides only 5.39$\times$ improvement in lifetime and 1.24$\times$ improvement in performance. Further, it incurs a loss of 4.94\% in energy. Our experiments have  also shown that the hybrid cache design provides better results than an SRAM-only cache of equal area on all the relevant evaluation metrics.

The rest of the paper is structured as follows.  Section \ref{sec:pcmbackground} discusses related work on cache design using NVM devices. Section \ref{sec:pcmmethods} discusses the design of hybrid cache along with DFB cache replacement policy. It also discusses the features, limitations and overhead of the DFB and compares it with the related work. Section \ref{sec:pcmexerimentalmethodology} discusses the simulation platform, energy model and evaluation metrics. Section \ref{sec:pcmresults} provides the results and finally, Section \ref{sec:pcmconclusion} presents the conclusion.





 \section{Related Work}\label{sec:pcmbackground}
Some researchers have proposed architectural techniques for managing power consumption of SRAM based last level caches \cite{MitZha12_EnCache,khaitan2013cache,masterMittal2013,mittalManager2013}, however, future extreme-scale computing systems and big-data processing in several fields \cite{agrawal2008new,raju2009high} present very strict demands on energy efficiency. These demands mandate fundamental changes at the device level and rethinking of the system architecture. For this reason, researchers have turned towards using NVMs for designing last level caches. 

Several researchers have proposed techniques to design NVM-only and SRAM-NVM hybrid cache designs. Joo et al. \cite{joo2010energy} propose several schemes for managing PCM-based caches. Their write minimization technique uses `read-before-write' operation to check whether the new value is different from the existing value and uses this observation to avoid redundant writes. To achieve wear-leveling, they use a bit-line shifter to spread out the writes over the whole PCM cells in a cache block.

Jadidi et al. \cite{jadidi2011high} present a hybrid cache architecture, where each cache set uses a few SRAM lines (cache ways) and a large number of STT-RAM lines. Their technique works by placing frequently-written data in SRAM ways and placing rarely-written/read-only data in STT-RAM ways. 

Since accesses to different sets are non-uniform, in a hybrid cache consisting of  SRAM and STT-RAM (or any other NVM) ways; the SRAM ways will be over-utilized in sets with high number of accesses and under-utilized in sets with small number of accesses.  Thus, the hybrid cache schemes cannot efficiently and symmetrically utilize the SRAM. Li et al. \cite{li2011exploiting} propose a technique which organizes the SRAM blocks in the hybrid cache as a semi-independent set-associative cache. Using this,  several hybrid cache sets can efficiently
share and cooperatively utilize their SRAM blocks, instead of exclusively utilizing the SRAM blocks in each cache set. Thus, the sets with over-utilized SRAM blocks can borrow SRAM space from other cache sets to properly accommodate the intensive write operations.

Some researchers have proposed use of embedded DRAM (eDRAM) for designing last level caches \cite{mittal2013edramcache}. However, eDRAM has retention period in the range of few microseconds and hence, eDRAM spends a significant fraction of energy in the form of refresh energy. This presents a crucial challenge in the use of eDRAM for designing LLCs.

In comparison to the above mentioned works, in this paper, we propose an SRAM-PCM hybrid cache design, and an intelligent cache replacement policy, named DFB (dead fast block) for managing this cache.






%
%

 \section{Methodology}\label{sec:pcmmethods}
Since PCM has high write latency/energy and small write endurance, a PCM-only cache is likely to have very small lifetime and also provide poor performance. For this reason, we propose a hybrid cache design, along with a cache replacement policy to manage this cache.
 
\subsection{Design of Hybrid Cache}
Our design of hybrid cache is based on several key observations and insights:

\begin{enumerate}
\item A hybrid cache should aim to achieve the best of both the worlds, viz. the high write endurance and speed of SRAM and high capacity (density) and low leakage power of PCM. Also, given the limited silicon chip-area, the cache should have similar or smaller area than an SRAM-only cache.
\item SRAM-PCM regions could be either across the sets or across the cache ways. Since using way-based hybrid cache design provides flexibility to use either of the devices with intelligent cache management policies, in this paper, we explore a way-based design, where few cache ways are designed using SRAM, while others are designed using PCM. 

\item With LRU-based cache replacement policies, most cache accesses are likely to hit in (or near) MRU way(s). In other words, few cache ways absorb most of the cache accesses, e.g. 2 out of 8 ways may absorb 80\% of the cache accesses. Hence, only few SRAM ways are sufficient.

\item L2 cache exhibits poorer locality than the L1 cache. Hence, using only a single SRAM way may not be sufficient to absorb most cache accesses. At the same time, using a large number of SRAM ways may increase the area and leakage power.

\end{enumerate}

We term the SRAM region as the \textit{fast} region and the PCM region as the \textit{slow} region. In what follows, we use the terms fast region and SRAM region interchangeably. Similarly, we use the terms slow region and PCM region interchangeably. 

The design of hybrid cache is parametrized over $N_{fast}$, which shows the number of ways which are designed using SRAM. Extreme values of $N_{fast}$ viz. 0 and $Assoc$ (which shows the cache associativity and is taken as 8 inn this work) give pure-PCM and pure-SRAM caches, respectively. Considering the trade-off between performance and energy efficiency, in this work, we take $N_{fast}=2$. We assume that the first $N_{fast}$ physical ways are designed using SRAM, although any other ways could be designed using SRAM. The physical design of hybrid cache can be done, as shown in previous works \cite{wu2009hybrid}.

\subsection{Design of DFB (Dead Fast Block) Replacement Policy}
We first explain the reasoning of how LRU replacement policy works and then explain the DFB policy. LRU tries to exploit the locality property of cache access to make a guess about the most suitable candidate for replacement. It records the  information about relative recency value of different blocks in a set, which is referred to as the LRU stack, the topmost value being the most recent. Whenever a block is accessed, it is promoted to the top of the stack. If a block reaches the bottom of the stack, it is understood that it has not been accessed recently, and hence, is not (or least) likely to be accessed again in the near future and hence, it can be evicted. Algorithm \ref{algo:lrurepl} shows the working of LRU replacement policy.

\begin{algorithm}
\caption{LRU (least recently used) cache replacement policy}\label{algo:lrurepl}
 \begin{algorithmic}[1]
\INPUT $LRU\_Order[1:Assoc]$ for all ways
\OUTPUT Index of the replacement candidate  
  \FOR{ $w$ = $1$ to $Assoc$}
     \IF  {$LRU\_Order[w] == (Assoc)$}
        \STATE RETURN $w$
     \ENDIF
\ENDFOR
 \end{algorithmic}
\end{algorithm}
Here, we have assumed that a larger value of $LRU\_Order$ denotes a least recently used (i.e. bottom of the LRU stack) block and vice-versa. Also, it is assumed that invalid block existing in a set (if any) have the largest $LRU\_Order$ value.

DFB further builds on the intuition that the blocks which are low in LRU stack are not likely to be used again and are soon going to be evicted. Thus, \textit{if such blocks are fast (SRAM) blocks, they can be evicted much early before they reach the bottom of the stack}. This provides the opportunity to reuse the fast block for storing the newly arrived data. Thus, a fast block is considered to be \textit{dead} much earlier than a slow block. For this reason, we term this replacement policy as \textit{dead fast block (DFB)} replacement policy. DFB uses a parameter $Z$ which shows the lower limit up to which a fast block is allowed to go (sink) in the LRU stack.  The working of DFB is shown in Algorithm \ref{algo:dfbrepl}. Here, we have assumed that at the beginning of program execution, the SRAM ways exist at the top of the LRU stack.

\begin{algorithm}
\caption{DFB (dead fast block) cache replacement policy}\label{algo:dfbrepl}
 \begin{algorithmic}[1]
\INPUT $LRU\_Order[1:Assoc]$ for all ways
\OUTPUT Index of the replacement candidate 
 \FOR{ $w$ = $1$ to $Assoc$}
   \IF{$w \leq N_{fast}$ and $LRU\_Order[w] \geq Z $}
   \STATE RETURN $w$
     \ELSIF  {$LRU\_Order[w] == (Assoc)$}
        \STATE RETURN $w$
     \ENDIF 
     \ENDFOR
 \end{algorithmic}
\end{algorithm}

Periodically, the value of $Z$ can be updated based on the locality present in the application cache access behavior. While more accurate and complicated methods of determining locality exist, in this work, we use a simple approach for determining cache access locality. Intuitively, applications with very high miss-rate have poor locality and vice-versa. Based on this and assuming $N_{fast}=2$ and $Assoc=8$, we use the following simple Algorithm (\ref{algo:dfbreplzupdate}) to update the value of $Z$. This algorithm runs after a fixed time period, for example, 5M cycles.

\begin{algorithm}
\caption{Algorithm for updating the value of $Z$ in DFB} \label{algo:dfbreplzupdate}
 \begin{algorithmic}[1]
\INPUT Cache miss-rate $M_r$(number of misses/number of accesses) in the last interval
\OUTPUT Updated value of $Z$  
     \IF     {$M_r < 80\%$  }
     \STATE $Z=5$
     \ELSIF  {$M_r < 90\%$}
     \STATE $Z=4$
     \ELSIF  {$M_r < 99\%$}
     \STATE $Z=3$
     \ELSE
     \STATE $Z=2$
     \ENDIF 
     \STATE RETURN $Z$     
 \end{algorithmic}
\end{algorithm}

\textbf{Features and Limitations of DFB:} The advantage of DFB comes from two reasons. First, since  the write endurance of SRAM is very high, for practical purposes, it is considered ``infinite''. DFB aims to bring SRAM way to the MRU position quickly and thus, it aims to direct future cache accesses to SRAM ways, which can sustain much larger number of writes. Also, the accesses to SRAM are much faster than those to PCM, and hence, by directing the cache accesses to SRAM ways, most of the writes are absorbed in fast SRAM ways. This leads to reduction in cache accesses to PCM ways, which reflects in improved performance, energy efficiency and lifetime. 

The limitation of DFB is that for applications with large working set and high access locality, DFB may lead to premature eviction of useful data blocks and may avoid evicting actually useless blocks which exist at the bottom of the LRU stack.

\textbf{Overhead of DFB:} It is clear that compared to LRU, DFB incurs minimal additional overhead. The value of $Z$ is same throughout the cache and its value is changed only after a large interval. For an 8-way cache, only 3 bits are required for storing $Z$. The value of $N_{fast}$ is fixed at the design time. DFB requires simple comparison operations and does not require prediction of future, storing additional data or complicated operations.   As we show in the results section, use of DFB provides significant improvement in energy efficiency, performance and lifetime of the hybrid cache.

\textbf{Comparison of DFB with Related Work:} Some researchers propose replacement policies for managing last level caches. Ferreira et al. \cite{ferreira2010increasing} propose a replacement policy for LLC which  aims to save PCM main memory energy by  reducing the write-back traffic to main memory. This policy is called $N$-Chance where $N$ can be varied. This policy evicts the least recently accessed clean block from the cache, unless all of the $N$ least recently accessed block are dirty, if so, it evicts the least recently accessed block. For the case when $N =1 $, this policy becomes the LRU policy. Their policy is different in that it is used in LLC for managing main memory designed using NVM, while in our case, the replacement policy is designed for LLC which is itself designed using NVM. Also, this policy exploits the difference between eviction of clean or dirty blocks, while DFB exploits the difference between access to fast or slow regions.

In the context of SRAM-NVM hybrid cache, some researchers have proposed cache management schemes for increasing the utilization of SRAM and mitigating write endurance problem of NVM. Wu et al. \cite{wu2009hybrid} propose an SRAM-NVM hybrid cache design, along with cache replacement and data migration policies to manage this cache.  Their technique records the  frequency of accesses to the lines in slow region and when it exceeds a threshold, the line is swapped with another line in the fast region. Other hybrid cache designs aim to place write-intensive blocks in fast region. Our approach is different in that, it does not uses migration/swapping which may require special hardware and create data coherence issues. Also, in the worst-case, migration/swap operations may themselves increase the write traffic significantly. We enforce the SRAM/PCM differentiation at the time of cache replacement. 

For L1 caches, Valero et al. \cite{valero2009hybrid}  present an SRAM-eDRAM cache design which uses 1 SRAM way and remaining eDRAM ways. Their technique aims to always keep the MRU way in the SRAM using a swap operation. However, since L2 shows reduced locality, this scheme is not effective for L2 cache and frequent swapping operations may incur large overhead. 

Some replacement policies such as pseudo-LRU, random, round-robin and several other policies \cite{al2004performance,kim2001lrfu,chaudhuri2009pseudo} aim to mitigate the implementation overhead of LRU or exploit the cache access behavior in a better manner to retain the useful data for a longer time. However, unlike DFB policy, they do not exploit the \textit{physical difference between cache ways}.



 \section{Evaluation Methodology}\label{sec:pcmexerimentalmethodology}
\subsection{Simulation Infrastructure}
We perform micro-architectural simulations using Sniper simulator with a processor frequency of  2GHz. All caches use a block size of 64 bytes. L1 I/D caches are 4-way LRU caches with 32KB capacity and 1 cycle latency. The details of L2 cache are provided below.  The memory queue has a bandwidth of 10 GB/s and queue contention is also modeled. The latency of main memory is 360 cycles. We use all SPEC2006 benchmarks and execute them for 250M instructions, after fast-forwarding them for 10B instructions.

\subsection{Parameters for SRAM, PCM and Hybrid Caches}
Table \ref{tab:hybridpcmsramvalues} shows the area, energy and timing values for a 1MB SRAM and an 8MB PCM cache which are obtained using NVsim \cite{dong2012nvsim}.  For computing these values, we have assumed a sequential cache access design, 350K temperature, H-tree organization of mats, 32nm CMOS process, 8-way associativity and 64B data block size. We find the designs which are optimized for write EDP.

\begin{table}[htbp]
  \centering
  \caption{Area, timing and energy values for SRAM and PCM}
    \begin{tabular}{|l|c|c|}
    \hline
          & 1MB SRAM & 8MB PCM \\    \hline
    Area (mm$^\text{2}$) & 1.894 & 1.602 \\\hline
    Hit Latency (ns) & 0.697 & 0.905 \\\hline
    Miss Latency (ns) & 0.217 & 0.274 \\\hline
    Write Latency (ns) & 0.3   & 150.384 \\\hline
    Hit Energy (nJ) & 0.29  & 3.326 \\\hline
    Miss Energy (nJ) & 0.006 & 0.969 \\\hline
    Write Energy (nJ) & 0.282 & 76.418 \\\hline
    Leakage Power (W) & 2.194 & 1.029 \\\hline
    
    \end{tabular}%
  \label{tab:hybridpcmsramvalues}%
\end{table}%

Taking a 1MB SRAM as the starting point, we look for iso-area design of a PCM cache and observe that an 8MB PCM has nearly same area as a 1MB SRAM. Therefore, we compare these two cache sizes. For designing the hybrid cache, we assume linear scaling of area values and observe that an SRAM-PCM hybrid cache where 2 ways are designed using SRAM and 6 ways are designed using PCM has an area of 1.675 mm$^\text{2}$, which is still smaller than the area of a 1MB SRAM cache, thus, under the condition of iso-area, this design is acceptable. We assume that the miss latency and miss energy of hybrid cache are same as that in a PCM cache. The hit and write energy/latency are incurred depending on whether the read or write is from an SRAM or a PCM region. The leakage power is assumed to scale linearly with the number of ways designed using SRAM and PCM, and thus, for the given hybrid cache, the leakage power is 1.320 Watts.

We evaluate the following designs
\begin{enumerate}
\item A PCM LLC with a capacity of 8MB, which uses LRU replacement policy.  This is our baseline.
\item An SRAM LLC with a capacity of 1MB, which uses LRU replacement policy. 
\item An SRAM-PCM hybrid cache design which uses LRU replacement policy and has a total capacity of 8MB. This is referred to as Hybrid-LRU design. 
\item An SRAM-PCM hybrid cache design which uses DFB replacement policy  and has a total capacity of 8MB. This is referred to as Hybrid-DFB design. 
\end{enumerate}
Both hybrid cache designs use 2 SRAM and 6 PCM ways. DFB uses an interval length of 5M cycles and at the beginning of program execution, $Z=4$. We assume that the 1MB SRAM cache uses a single-bank structure and hence, allows only one ongoing cache access at a time. Both PCM-only and hybrid caches use an 8-bank structure and hence, allow 8 concurrent cache accesses.

We model the energy consumed in L2 cache and main memory. The dynamic energy and leakage power of main memory are 70 nJ/access and 0.18 Watt, respectively \cite{MittalPalettePaper2013,MitZha13_Cashier}. The L2 cache energy is computed using the values from Table \ref{tab:hybridpcmsramvalues}.

\subsection{Evaluation Metrics}

We show the results on percentage saving in energy, relative performance (i.e., ratio of IPC for a scheme and IPC for the baseline), and absolute increase in MPKI (miss-per-kilo-instructions). Further, we show the relative lifetime, where the lifetime is defined as the inverse of maximum number of writes on any cache block. The value of relative performance and relative lifetime are summarized using geometric mean, while the remaining metrics are summarized using arithmetic mean. 


\section{Results}\label{sec:pcmresults}
\begin{figure*}[htbp]
 \centering
 \includegraphics [scale=0.5] {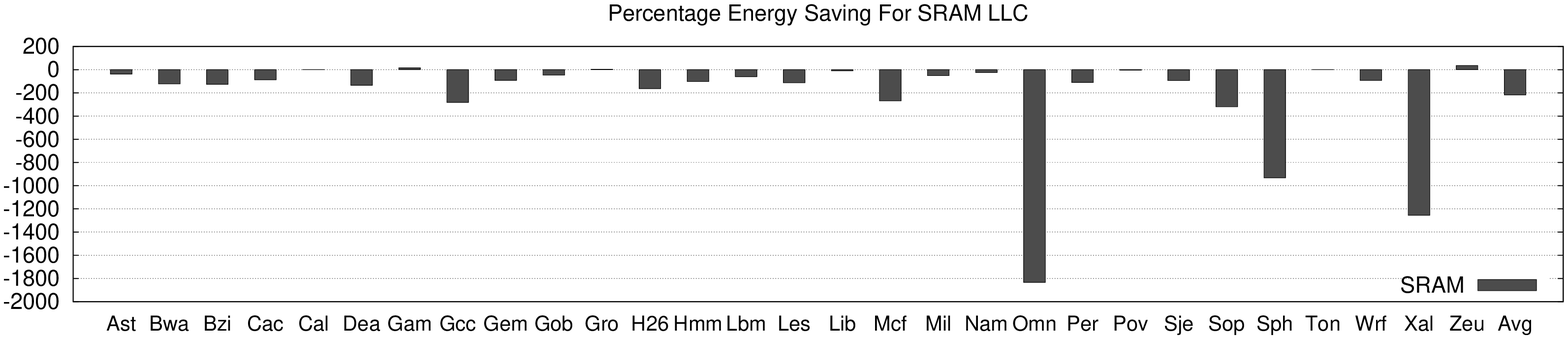} 
 \includegraphics [scale=0.5] {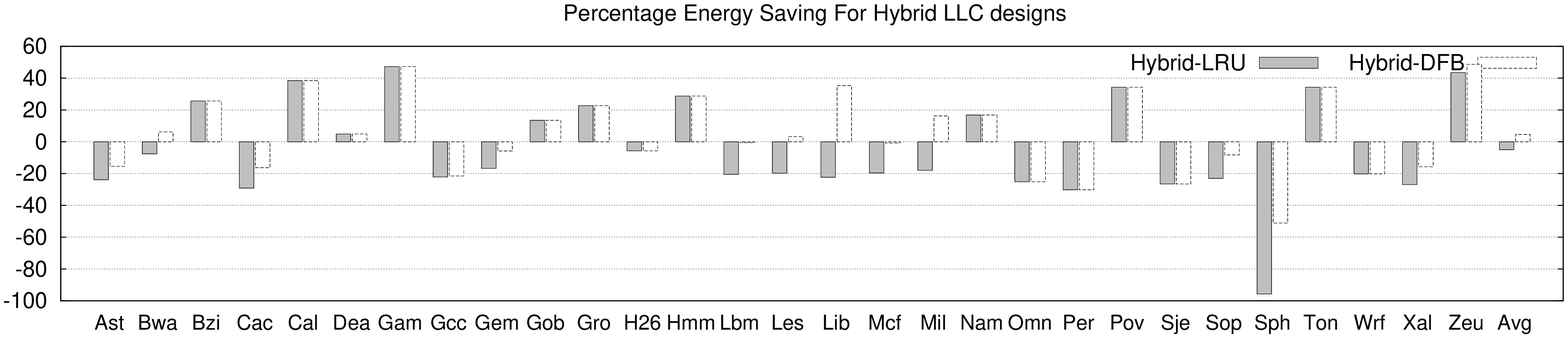}
 \includegraphics [scale=0.5] {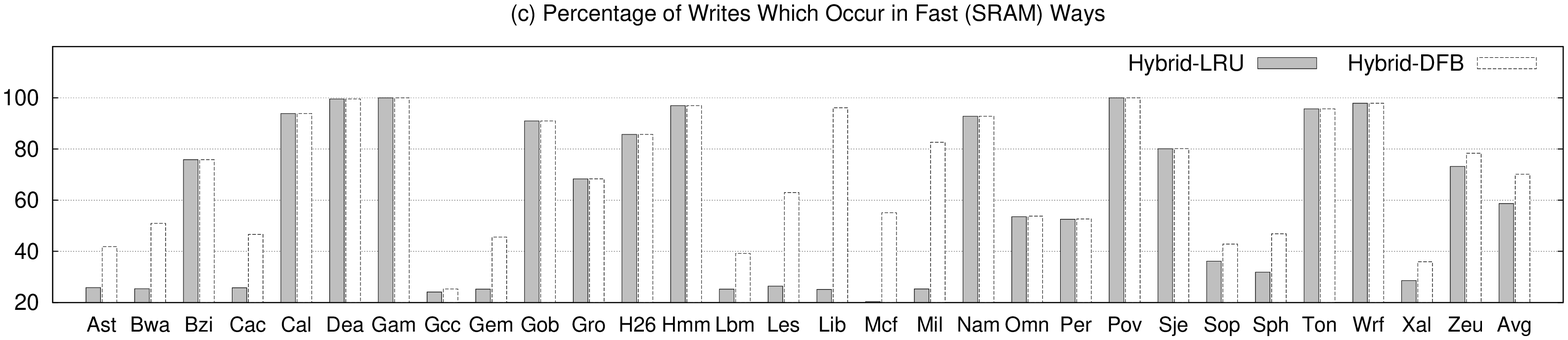}
 \includegraphics [scale=0.5] {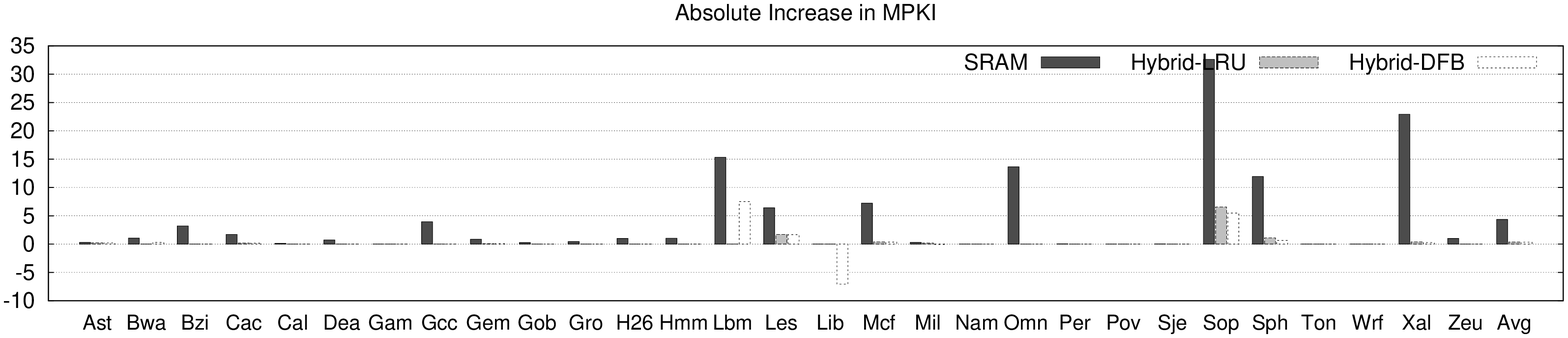}
 \includegraphics [scale=0.5] {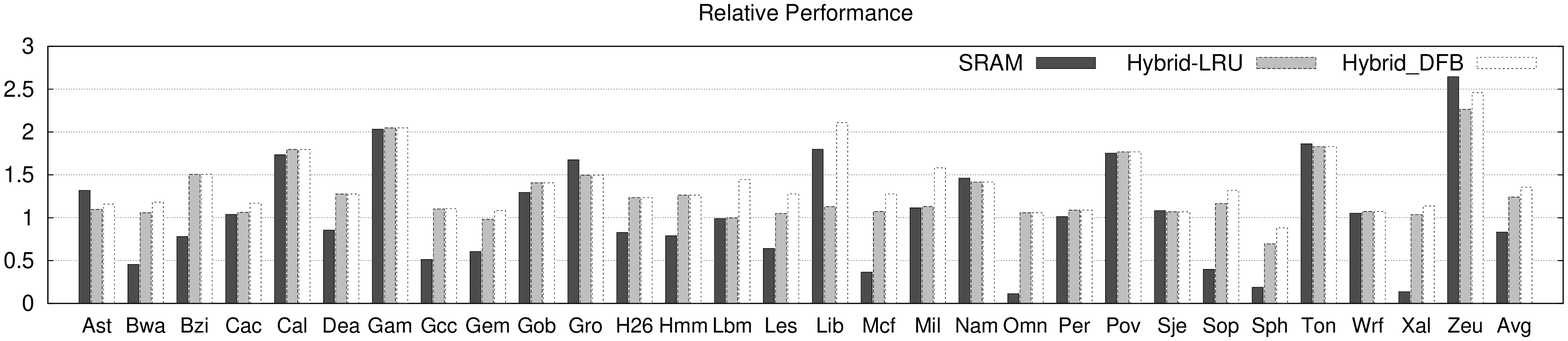} 
 \caption{Results on percentage energy saving, fraction of writes in fast ways, absolute increase in MPKI and relative performance}
\label{fig:hybridpcmresultsenergysaving}
 \end{figure*}

\begin{figure*}[htbp]
 \centering
 \includegraphics [scale=0.5] {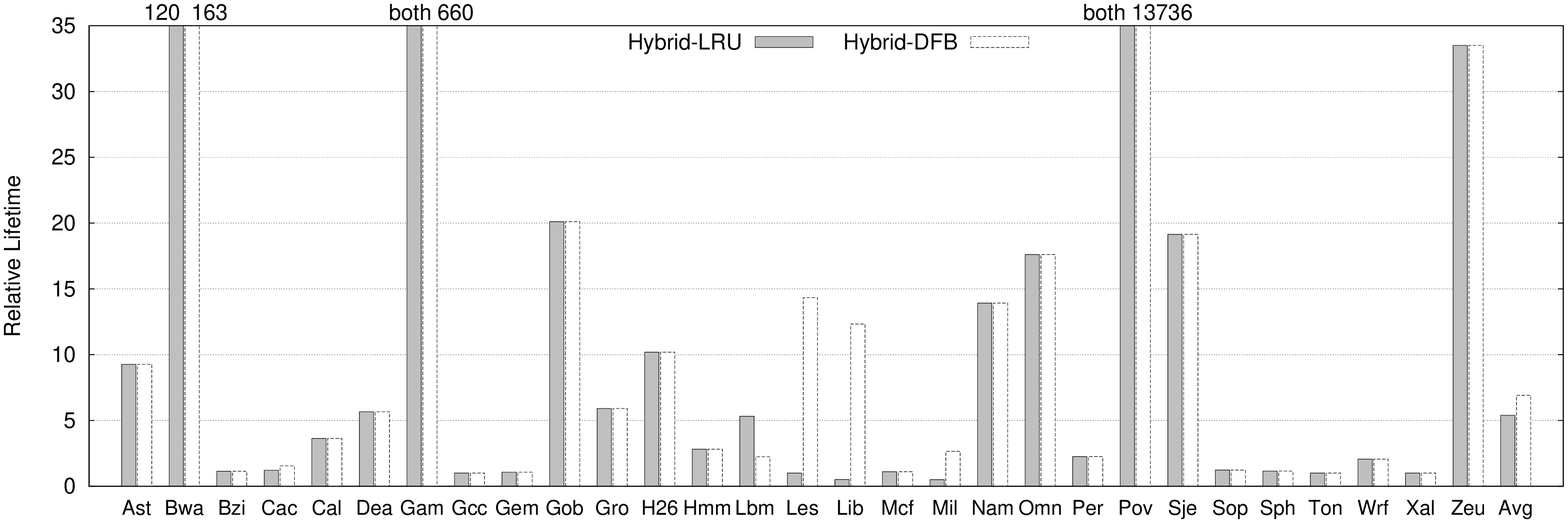} 
 \caption{Results on relative lifetime}
\label{fig:hybridpcmresultslifetime}
 \end{figure*}

Figure \ref{fig:hybridpcmresultsenergysaving} shows the results on percentage energy saving, fraction of writes in fast ways, MPKI increase and relative performance. Since the range of values of energy saving are different for SRAM and Hybrid caches, we show them in separate figures. Note that a negative value of energy saving implies loss in energy.  Figure \ref{fig:hybridpcmresultslifetime} shows the results on relative lifetime.

We now analyze the results in detail. For applications with small working set (e.g. povray, perlbench, sjeng, namd, tonto, gamess etc.), the extra capacity provided by PCM cache does not help in reducing the miss-rate. On the other hand, applications with large working set (e.g. soplex, xalancbmk, lbm, omnetpp etc.) benefit significantly from the extra capacity and hence, their miss-rate is significantly reduced on using PCM or hybrid cache. 

Compared to the PCM cache, the Hybrid-LRU cache shows poor energy efficiency, despite the fact that SRAM has faster and more energy efficient writes. This is because, SRAM also has larger leakage power consumption which forms a significant fraction of energy consumption in the LLCs. For some applications, SRAM cache incurs very high energy loss, for example, for omnetpp, the SRAM LLC incurs more than 1800\% loss in energy and for xalancbmk, the SRAM LLC incurs more than 1200\% loss in energy. This clearly shows that despite their high write latency/energy, NVM caches can provide advantage over conventional SRAM caches. Since energy efficiency is now becoming the primary constraint in processor design \cite{Mit_DRAMsurvey}, NVM devices provide attractive alternative to conventional devices for fabricating  processor components, such as caches.  

As for performance, on average, SRAM cache shows poor performance than the PCM cache, since the lower capacity of SRAM makes its miss-rate very high for cache intensive applications. As discussed above, for applications with small working set size, SRAM provides better performance, since for the same miss-rate, the faster read/write speed of SRAM helps in reducing the latency.

For the applications, where a large number of writes occur in SRAM ways, the lifetime is greatly increased since the number of writes to PCM is significantly reduced. The most dramatic improvement is seen in povray, where the improvement in lifetime is 13736$\times$. This can be easily explained from Figure \ref{fig:hybridpcmresultsenergysaving}(c), where we can see that for povray, nearly 100\% of the writes are absorbed by the SRAM ways. Also, povray has highly non-uniform writes to different cache sets and thus, only few blocks are heavily written, which in the case of hybrid cache are the SRAM ways. Similarly, a high increase in lifetime of other applications such as bwaves, gamess, zeusmp can be explained. On average, the improvement in lifetime using Hybrid-LRU and Hybrid-DFB are 5.39$\times$ and 6.91$\times$. Thus, compared to a PCM-only cache, the Hybrid-DFB cache design provides nearly 7$\times$ large cache lifetime. The larger improvement in lifetime on using Hybrid-DFB compared to that from Hybrid-LRU can be explained based on the reasoning above and Figure \ref{fig:hybridpcmresultsenergysaving}(c). 

An interesting behavior of DFB replacement policy is seen in the case of libquantum benchmark. With LRU policy, libquantum is a streaming benchmark with a miss-rate of 100\%. However, with DFB, its miss-rate is less than 100\%. The reason for this is that DFB confines the replacement activities within a set to the upper (top) part of the LRU-stack and thus, leaves some blocks in the lower part, which are not immediately evicted. These blocks are later reused and hence, they see cache hit. LRU replacement policy, however, does not leverage this advantage.

For all the metrics shown here, viz. energy saving, performance improvement, lifetime improvement, writes to fast ways and increase in MPKI, Hybrid-DFB performs better than the Hybrid-LRU design. Specifically note that while Hybrid-LRU incurs nearly 4.9\% loss in energy over a PCM baseline, the Hybrid-DFB achieves nearly 4.5\% saving in energy. This shows the effectiveness of our approach.

\section{Conclusion}\label{sec:pcmconclusion}
In this paper, we have presented a hybrid cache designed using SRAM-PCM, which aims to utilize the best features from both the devices. We also presented a cache replacement policy for managing this cache which aims to increase the number of writes to SRAM regions and thus, improves the cache lifetime and energy efficiency. Experiment results have been presented using SPEC2006 benchmarks and the Hybrid-DFB cache design has been compared with SRAM cache, pure-PCM cache and Hybrid-LRU cache design. The results show that, our approach provides the best results on all the evaluation metrics.




\ifCLASSOPTIONcaptionsoff
  \newpage
\fi



\bibliographystyle{IEEEtran}
\bibliography{PhDReferences}

\end{document}